\documentclass[journal]{IEEEtran}
\usepackage{amssymb}
\usepackage{amsmath}
\usepackage{graphicx}
\usepackage{cite}
\usepackage{balance}
\usepackage[utf8x]{inputenc}
\usepackage{fancyhdr}
\usepackage{amsmath,amssymb,amsthm}
\usepackage{graphicx,epstopdf}
\usepackage{epsfig}	
\usepackage{subfigure}
\usepackage{lipsum}
\usepackage{bbm}
\usepackage{stfloats}
\usepackage{xcolor}
\usepackage{textcomp}
\usepackage{algorithm}
\usepackage{algorithmic} 
\usepackage{makecell}
\usepackage{multirow}

\DeclareMathAlphabet{\mathpzc}{OT1}{pzc}{m}{it}

\newtheorem{theorem}{Theorem}

\newtheorem{lemma}{Lemma}

\newtheorem{corollary}{Corollary}

\begin{document}
\title{A Model-Driven Deep Learning Method for Massive MIMO Detection}
\author{Jieyu~Liao, Junhui~Zhao, Feifei~Gao and Geoffrey~Ye~Li
%
%
}
\maketitle

\begin{abstract}
In this paper, an efficient massive multiple-input multiple-output (MIMO) detector is proposed by employing a deep neural network (DNN).
Specifically, we first unfold an existing iterative detection algorithm into the DNN structure, such that the detection task can be implemented by deep learning (DL) approach.
We then introduce two auxiliary parameters at each layer to better cancel multiuser interference (MUI).
The first parameter is to generate the residual error vector while the second one is to adjust the relationship among previous layers.
We further design the training procedure to optimize the auxiliary parameters with pre-processed inputs.
The so derived MIMO detector falls into the category of model-driven DL.
The simulation results show that the proposed MIMO detector can achieve preferable detection performance compared to the existing detectors for massive MIMO systems.
\end{abstract}

\IEEEpeerreviewmaketitle
\section{Introduction}
Massive multiple-input multiple-output (MIMO) is regarded as a promising technology to achieve higher spectral and power efficiency in current wireless communication systems \cite{Zhao2019Pilot}, \cite{Wang2019Spatial}.
Since the transmitters and receivers are equipped with tens or hundreds of antennas, the entire signal processing of MIMO becomes complicated, especially for MIMO detection.
For example, the inter-user interference \cite{Zhao2019Multiband} significantly affects the detection accuracy,
and the multiple data sequences to be detected prolong the delay of the communication systems.

The maximum likelihood (ML) detector is optimal but can only be solved by ``brute-force" search, whose computational complexity increases exponentially with the number of antennas \cite{Yang2015Fifty}.
Consequently, the near-optimal detectors that could provide acceptable performance with low complexity are preferable \cite{wei2019learned},
e.g., the approximate message passing (AMP) detector\cite{Borgerding2017AMP-Inspired} and the semidefinite relaxation (SDR) detector\cite{Luo2010Semidefinite}, etc.
However, the complexity of AMP increases with the number of users and the order of modulation\cite{Zeng2018Low} and SDR is limited in the constellations \cite{samuel2019Learning}.
In terms of massive MIMO scenarios,
linear detectors themselves, such as zero forcing (ZF) detector\cite{Yang2015Fifty} and the linear minimum mean-squared error (LMMSE) detector\cite{Yang2015Fifty} are with low complexity, but finding the coefficients of the detectors usually needs matrix inversion and is complicated.

Recently, deep learning (DL) has made plausible success in many fields, such as image recognition, nature language processing and speech recognition.
In fact, the application of DL in wireless communication systems can be classified into two categories \cite{Qin2019Deeplearning}: data-driven method and model-driven method \cite{He2019Model-Driven}.
Data-driven method learns the characteristic directly from a large number of data, and has been applied in channel estimation \cite{Yang2019DeepLearning-Based}, \cite{Ye2018powerofdeeplearning},
CSI feedback \cite{Wen2018CSIfeedback} and MIMO detection \cite{samuel2017Deep}, \cite{samuel2019Learning}. Take MIMO detection as example:
an algorithm, named DetNet, in \cite{samuel2017Deep} combines fully connected neural network and MIMO detector.
DetNet is formulated by unfolding a projected gradient descent algorithm for ML optimization,
and exhibits better performance than AMP and SDR detectors at the expense of offline training process for tens of hours.
Although the data-driven method achieves success and has been studied well, the feature of learning from data
requires a large sample set and is time consuming.
In addition, the model-driven approach can mitigate the time and sample consumption issues.
Model-driven method \cite{He2019Model-Driven} optimizes the parameters or add some parameters learned by DL in the existing model,
which combines the advantages of data-driven method and conventional mathematical models.
For example,
the algorithm introduced in \cite{He2018AModel-Driven} adopts DL to optimize the parameters in orthogonal AMP (OAMP) model,
which improves the performance of the classical OAMP algorithm in term of bit-error rate (BER).


In this paper,
a new model-driven DL-based massive MIMO detector is proposed by trickily unfolding an existing iterative algorithm \cite{Mandloi2017Low-Complexity} for the multiuser interference cancellation.
We use auxiliary parameters to involve the previous residual vector and design training procedure.
The simulation results show that DNN has the ability to learn and analyse the characteristics of iterative architectures to adapt to the network for lower BER performance.

\section{DL for MIMO Detection}
\subsection{System Model}
Consider a system where base station (BS) is equipped with $N$ antennas and $K$ users are equipped with single antenna.
In the uplink transmission, the received signal at the BS can be expressed as
\begin{equation}\label{eq:1}
{\bf{\widetilde{y} = \widetilde{H}\widetilde{x} + \widetilde{n}}},
\end{equation}
where $\mathbf{\widetilde{x}}$ denotes the transmitted symbol vector drawn from the constellation alphabet $\mathbb{A}$, $\mathbf{\widetilde{H}}$ denotes the channel matrix and $\mathbf{\widetilde{n}}$ is the additive white Gaussian noise (AWGN) that is generated from $\mathcal{CN}(0, {\sigma ^2}{{\rm \mathbf{I}}_N})$.

To generalize the aforementioned model, we transform \eqref{eq:1} into real domain as
\begin{equation}\label{eq:2}
\mathbf{y} = \mathbf{H}\mathbf{x} + \mathbf{n},
\end{equation}
where
\[{\bf{H}} \buildrel \Delta \over = \left[ {\begin{array}{*{20}{c}}
{\Re \left( {{\bf{\tilde H}}} \right)}&{ - \Im \left( {{\bf{\tilde H}}} \right)}\\
{\Im \left( {{\bf{\tilde H}}} \right)}&{\Re \left( {{\bf{\tilde H}}} \right)}
\end{array}} \right],\]
\[{\bf{y}} \buildrel \Delta \over = \left[ {\begin{array}{*{20}{c}}
{\Re \left( {{\bf{\tilde y}}} \right)}\\
{\Im \left( {{\bf{\tilde y}}} \right)}
\end{array}} \right],
{\bf{x}} \buildrel \Delta \over = \left[ {\begin{array}{*{20}{c}}
{\Re \left( {{\bf{\tilde x}}} \right)}\\
{\Im \left( {{\bf{\tilde x}}} \right)}
\end{array}} \right],
{\bf{n}} \buildrel \Delta \over = \left[ {\begin{array}{*{20}{c}}
{\Re \left( {{\bf{\tilde n}}} \right)}\\
{\Im \left( {{\bf{\tilde n}}} \right)}
\end{array}} \right],\]
and $\Re \left(  \cdot  \right)$ and $\Im \left(  \cdot  \right)$ denote the real and imaginary parts, respectively, ${\bf{H}} \in {\mathbb{R}^{2N \times 2K}}$, ${\bf{y}} \in {\mathbb{R}^{2N}}$, ${\bf{w}} \in {\mathbb{R}^{2N}}$ and ${\bf{x}} \in {\mathbb{R}^{2K}}$.

The detected signal vector for the ZF detector is given by \cite{Yang2015Fifty}
\begin{equation}\label{eq:3}
{{\bf{\hat x}}_{ZF}} = \mathcal{Q}[{{\bf{(}}{{\bf{H}}^{T}}{\bf{H)}}^{{ - 1}}}{{\bf{H}}^{T}}{\bf{y}}],
\end{equation}
where $\mathcal{Q}[\cdot]$ is the quantizer. 
Even if the linear detector itself is with low complexity in uplink massive MIMO systems, the inverse operation in \eqref{eq:3} to find the coefficient of the detector has complexity as high as $\mathcal{O}({K^3})$.

\subsection{Deep Learning Methods}
The basic structure of a DNN is shown in Fig. 1.
\begin{figure}[!t]
  \centering
  \includegraphics[width=8.7cm]{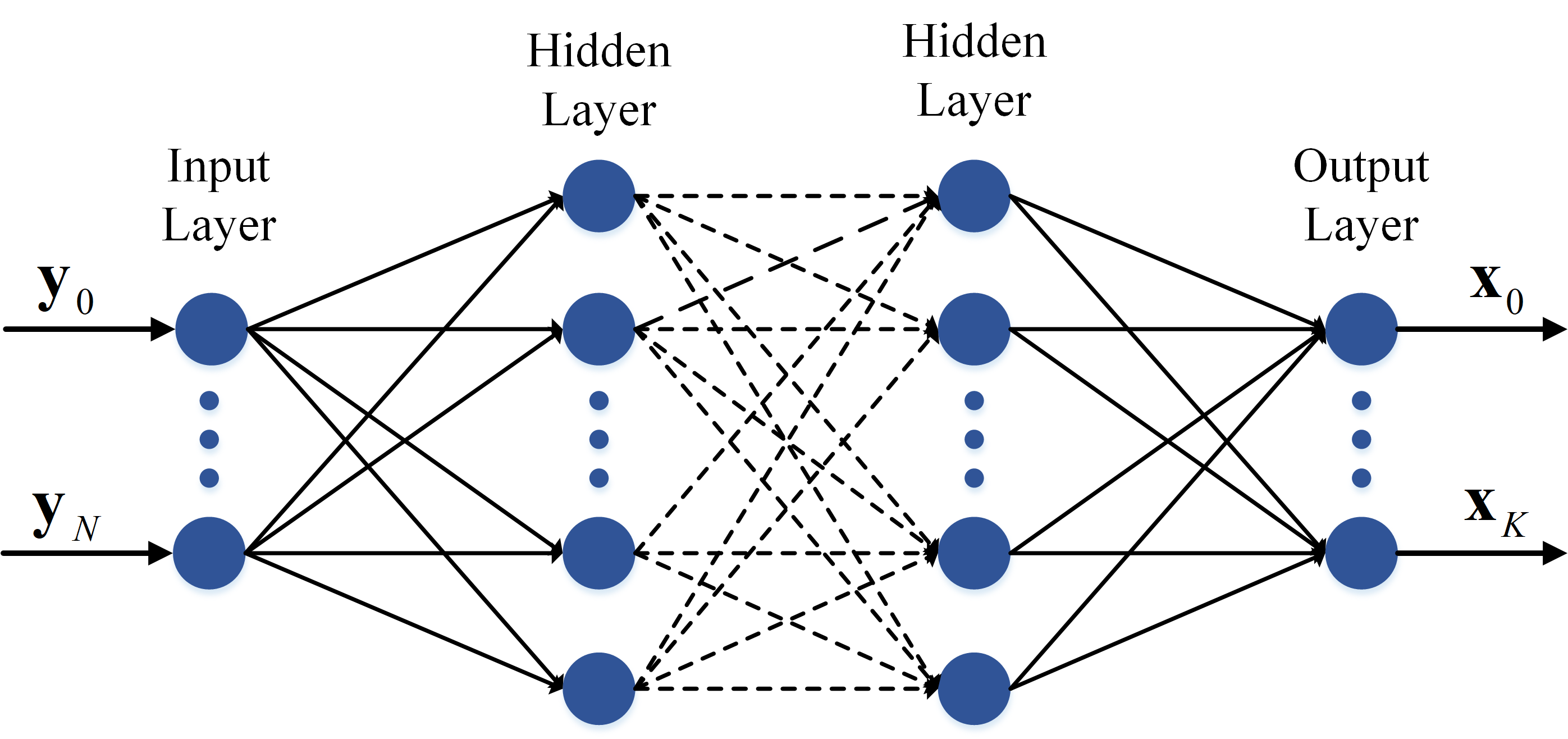}\\  
  \caption{The structure of the DNN}
\end{figure}
Generally, it consists of input layer, hidden layers, and output layer.
Moreover,
a number of activation functions can be adopted as the leaky rectified linear unit (Leaky ReLu) or the tan hyperbolic (Tanh) function,
defined respectively as
\begin{equation}\label{eq:15}
{f_L}\left( x \right) = \left\{ {\begin{array}{*{20}{c}}
{\!\!\!x,\;{\rm{  if\; }}x \ge 0{\rm{ }}}\\
{\frac{x}{a}{\rm{,\; if \;}}x < 0}\;,
\end{array}} \right.
\end{equation}
\begin{equation}\label{eq:16}
{f_T}\left( x \right) = \frac{{{e^x} - {e^{ - x}}}}{{{e^x} + {e^{ - x}}}}.\;\;\;\;\;\;\;\;\;\;
\end{equation}

Consequently, the output of the network is
\begin{equation}\label{eq:12}
\mathbf{o} = f\left( {\mathbf{z},\bf{\Theta }  } \right) = {f^{\left( L \right)}}\left( {{f^{\left( {L - 1} \right)}}\left( { \ldots {f^{\left( 1 \right)}}\left( \mathbf{z} \right)} \right)} \right),
\end{equation}
where $L$ stands for the number of layers, ${\bf{\Theta }}$ denotes the learned parametric set (also denotes weights), and $\mathbf{z}$ is the input of the network.
The parametric set ${\bf{\Theta }}$ is optimized by reducing the loss defined as the distance between the prediction and the regression vector.
In DL, we define $D$ as the size of the dataset. The pair $\left\{ {\left( {{\mathbf{z}^{\left( d \right)}},{\mathbf{x}^{\left( d \right)}}} \right)} \right\}_{d = 1}^D$ is utilized to train ${\bf{\Theta }}$. Then, the obtained ${\bf{\Theta }}$ is adopted to predict the regression vector $\mathbf{x}$ from a subset of feature $\mathbf{z}$.

\section{Model-Driven DL-based Detector}
In this section, we propose a model-driven DL-based algorithm for MIMO detection.
We first modify the existing algorithm to obtain the DNN structure and then develop the joint training procedure.
Finally, we compare the computational complexity with several existing algorithms.
\begin{figure}[!t]
  \centering
  \includegraphics[width=9cm]{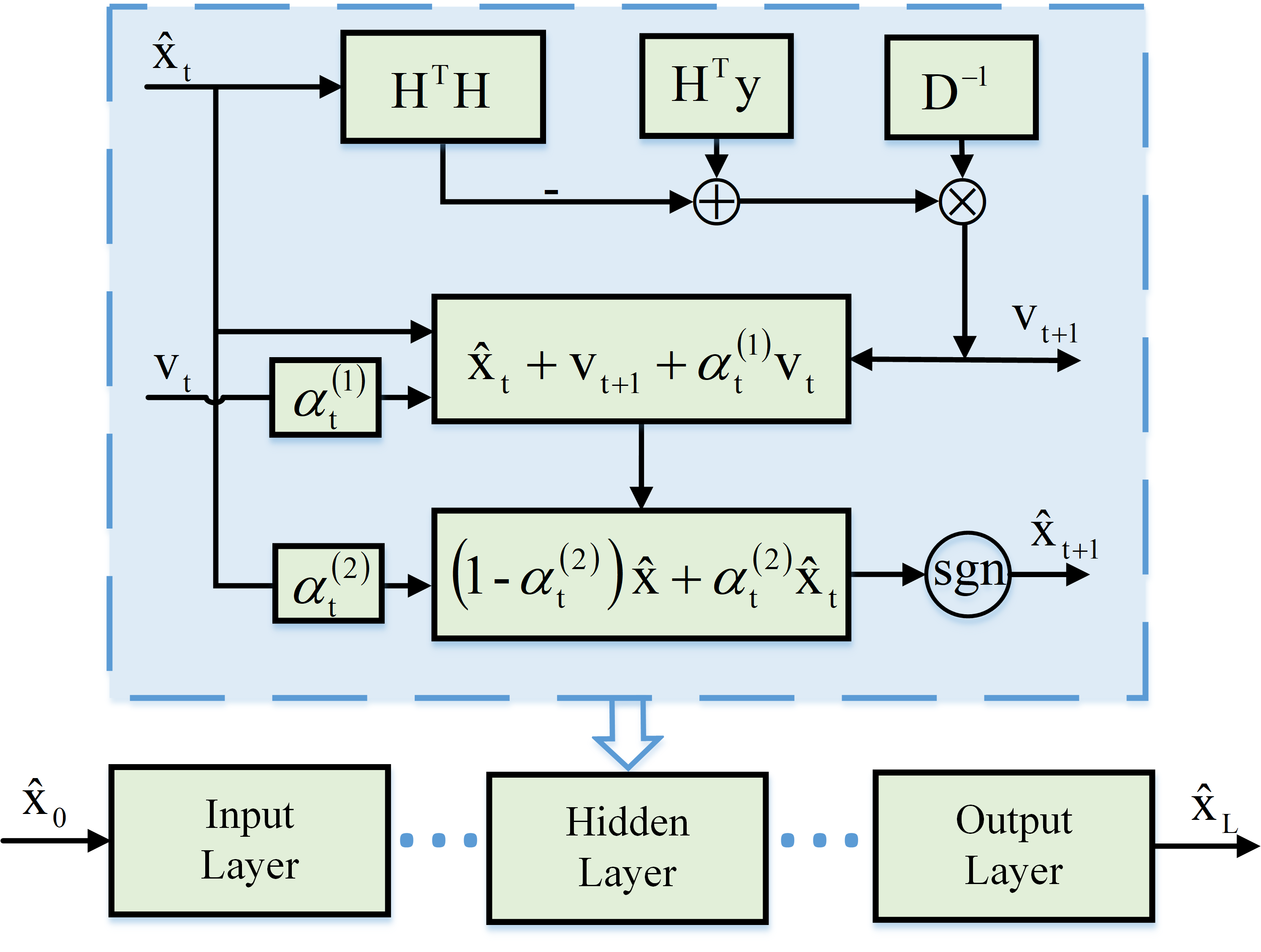}\\  
  \caption{A flowchart representing one layer of the proposed algorithm}
\end{figure}

\subsection{DNN Structure for Model-Driven DL-based MIMO Detection}
The flowchart in Fig. 2 presents each layer of the proposed DL-based detection network, which is originated from the work in \cite{Mandloi2017Low-Complexity}.
The iterative algorithm in \cite{Mandloi2017Low-Complexity} has $L$ iterations, each of which is mapped to a layer of the proposed DNN structure, i.e., the layer of DNN is $L$.


At the $t$-th layer of the DNN,
the input vector is already calculated from previous layer as
\begin{equation}\label{eq:8}
{{{\bf{\hat x}}}_t} = {[{{\hat x}_t}\left( 1 \right),{{\hat x}_t}\left( 2 \right), \cdots ,{{\hat x}_t}\left( {2K} \right)]^T}.
\end{equation}
The function of the $t$-th layer is to calculate ${{{\bf{\hat x}}}_{t+1}}$ from ${{{\bf{\hat x}}}_t}$, and pass it to the next layer.

To detect ${{{\bf{\hat x}}}_{t+1}(i)}$, the received signal that eliminates the interference from other users can be expressed as
\begin{equation}\label{eq:interference}
{\bf{\hat y}}_ i = {\bf{y}} - \sum\limits_{k = 1,k \ne i}^{2K} {{{\bf{h}}_k}} {{\hat x}_t}\left( k \right) = {{\bf{h}}_i}{{\hat x}_{t + 1}}\left( i \right),
\end{equation}
where ${{\bf{h}}_k}$ is the $k$-th column of $\mathbf{H}$.

Therefore, the output of the $t$-th layer is
\begin{equation}\label{eq:detect}
\begin{array}{l}
 {{\hat x}_{t + 1}}\left( i \right) = \frac{{{{\bf{h}}_i}^T}}{{{{\left\| {{{\bf{h}}_i}} \right\|}^2}}}{\bf{\hat y}}_i \\
  \;\;\;\;\;\;\;\;\;\;\;\;\;= \frac{1}{{{{\left\| {{{\bf{h}}_i}} \right\|}^2}}}\left( {{{\bf{h}}_i}^T{\bf{y}} - {{\bf{h}}_i}^T\sum\limits_{k = 1,k \ne i}^{2K} {{{\bf{h}}_k}} {{\hat x}_t}\left( k \right)} \right) \\
  \;\;\;\;\;\;\;\;\;\;\;\;\;= \frac{1}{{{{\left\| {{{\bf{h}}_i}} \right\|}^2}}}\!\left(\! {{{\bf{h}}_i}^T{\bf{y}}\! - \!{{\bf{h}}_i}^T\sum\limits_{k = 1}^{2K} {{{\bf{h}}_k}} {{\hat x}_t}\left( k \right)} \!\right)\! +\! \frac{{{{\bf{h}}_i}^T{{\bf{h}}_i}}}{{{{\left\| {{{\bf{h}}_i}} \right\|}^2}}}{{\hat x}_t}\left( i \right), \\
 \end{array}
\end{equation}
for $i = 1,\cdots,2K$.

Then, the detection task is transformed into the recursive form as
\begin{equation}\label{eq:recursion}
{{\hat x}_{t + 1}}\left( i \right) = {{\hat x}_t}\left( i \right) + \frac{1}{{{{\left\| {{{\bf{h}}_i}} \right\|}^2}}}\left( {{{\bf{h}}_i}^T{\bf{y}} - {{\bf{h}}_i}^T\sum\limits_{k = 1}^{2K} {{{\bf{h}}_k}} {{\hat x}_t}\left( k \right)} \right).
\end{equation}

Define a diagonal matrix $\mathbf{D}$ as
\begin{equation}\label{eq:diag}
{\bf{D}} = \mathrm{diag}\left( {{{\bf{H}}^T}{\bf{H}}} \right) = \mathrm{diag}\left\{ {{d_1},{d_{2,}} \cdots {d_i}, \cdots ,{d_k}} \right\},
\end{equation}
where ${d_i} = {\left\| {{{\bf{h}}_i}} \right\|^2}$.

The equation \eqref{eq:recursion} can be expressed as
\begin{equation}\label{eq:transform}
{{\hat x}_{t + 1}}\left( i \right) = {{\hat x}_t}\left( i \right) + \frac{1}{{{{d}_i}}}\left( {{{\bf{h}}_i}^T{\bf{y}} - \sum\limits_{k = 1}^{2K} {{{\left( {{{\bf{H}}^T}{\bf{H}}} \right)}_{i,k}}} {{\hat x}_t}\left( k \right)} \right).
\end{equation}
Equation \eqref{eq:transform} can be written in matrix-vector form as
\begin{equation}\label{eq:vector}
{{{\bf{\hat x}}}_{t + 1}} = {{{\bf{\hat x}}}_t} + {{\bf{v}}_{t+1}},
\end{equation}
where
 ${{\bf{v}}_{t+1}} = {{\bf{D}}^{ - 1}}\left( {{{\bf{H}}^T}{\bf{y}} - {{\bf{H}}^T}{\bf{H}}{{{\bf{\hat x}}}_t}} \right)$ is the residual error vector.

However, the above process of \eqref{eq:detect}-\eqref{eq:vector} could not ideally eliminate the interference and hence,
${{{\bf{\hat x}}}_{t + 1}}$ is influenced not only by ${{\bf{v}}_{t + 1}}$, but also by previous  ${{\bf{v}}_t},{{\bf{v}}_{t - 1}}, \cdots ,{{\bf{v}}_1}$ (note that ${{\bf{v}}_0}$ is self-defined, which has no obvious physical meaning).
Motivated by this,
we propose to detect ${{{\bf{\hat x}}}_{t + 1}}$ by
\begin{equation}\label{eq:model}
{{{\bf{\hat x}}}_{t + 1}} = {{{\bf{\hat x}}}_t} + {{\bf{v}}_{t{\rm{ + 1}}}}{\rm{ + }}\:\alpha _t^{\left( 1 \right)}{{\bf{v}}_t} + \alpha _{t - 1}^{\left( 1 \right)}{{\bf{v}}_{t - 1}} +  \cdots  + \alpha _1^{\left( 1 \right)}{{\bf{v}}_1},
\end{equation}
where $\alpha _t^{\left( 1 \right)}$, $\alpha _{t-1}^{\left( 1 \right)}$, $\cdots$, $\alpha _1^{\left( 1 \right)}$ are the parameters to be learned by DL.

Since the correlation between the adjacent residual vectors is the strongest, we here only consider the influence of ${{\bf{v}}_t}$ at the $t$-th layer for simplicity.
Then, equation \eqref{eq:model} is simplified to
\begin{equation}\label{eq:theta1}
{{{\bf{\hat x}}}_{t + 1}} = {{{\bf{\hat x}}}_t} + {\bf{ }}{{\bf{v}}_{t{\rm{ + 1}}}}{\rm{ +\; }}\alpha_t^{\left( 1 \right)}{{\bf{v}}_t}.
\end{equation}
Since ${{\bf{v}}_t}$ is involved in the calculation of ${{{\bf{\hat x}}}_{t + 1}}$, then ${{\bf{v}}_{t+1}}$ should be passed from the $t$-th layer to the $(t+1)$-th layer.
Hence, we need to modify the current DNN structure such that the input and output of DNN contain ${{\bf{v}}_t}$ and ${{\bf{v}}_{t+1}}$, respectively.

Since the value of equation \eqref{eq:theta1} is continuous, we regress the estimation to certain point as
\begin{equation}\label{eq:qantizex}
{{{\bf{\hat x}}}^q}_{t + 1} = \mathcal{Q}[{{{\bf{\hat x}}}_t} + {\bf{ }}{{\bf{v}}_{t{\rm{ + 1}}}}{\rm{ +\; }}\alpha_t^{\left( 1 \right)}{{\bf{v}}_t}],
\end{equation}
where $\mathcal{Q}[.]$ is the quantizer.

As in most iterative detection algorithm, we would set an upper bound for the number of layers in reality.
In this case, ${{{\bf{\hat x}}}^q}_{t + 1}$ may not converge within limited number of layers.

We adopt the convex combination of ${{{\bf{\hat x}}}^q}_t$ and ${{{\bf{\hat x}}}_{t + 1}}$,
defined as $\sum\limits_{i = t}^{t + 1} {\alpha _i^{\left( 2 \right)}} {{{\bf{\hat x}}}_i}$ with $\sum\limits_{i = t}^{t + 1} {\alpha _i^{\left( 2 \right)}}  = 1$.
Moreover, parameter ${\alpha ^{\left( 2 \right)}}$ will be optimized by DL.
Different from the residual error vector,
the optimization of ${\alpha ^{\left( 2 \right)}}$ at each layer would not change the DNN structure since the $t$-th layer has the knowledge of ${{{\bf{\hat x}}}^q}_t$.
For the implementation consideration, we choose ${{{\bf{\hat x}}}^q}_t$ to do the convex combination.
Then the outputs of each layer are ${{{\bf{\hat x}}}^q}$ and $\mathbf{v}$.

As a result, the detection of the $t$-th layer is
\begin{equation}\label{eq:qautizer}
{{{\bf{\hat x}}}^q}_{t + 1} = {\cal Q}[(1 - \alpha _t^{\left( 2 \right)}){{{\bf{\hat x}}}_{t + 1}} + \alpha _t^{\left( 2 \right)}{\bf{\hat x}}_t^q].
\end{equation}

Additionally, to achieve a lower BER at higher-order modulation situation, we have slightly modified the structure.
Since higher-order modulation needs more flexibility to obtain a lower BER,
we add two layers to modify $\mathbf{v}_t$ before multiplying $\alpha _t^{\left( 1 \right)}$ as
\begin{equation}\label{eq:16qam}
{\mathbf{v}_t}  \leftarrow  \mathbf{W}_t^2\left( {\mathbf{W}_t^1{\mathbf{v}_t} + \mathbf{b}_t^1} \right) + \mathbf{b}_t^2,
\end{equation}
where $\mathbf{W}_t^1$ ($\mathbf{b}_t^1$) and $\mathbf{W}_t^2$ ($\mathbf{b}_t^2$) are the first and the second weight (bias) at the $t$-th layer. There is no activation function here since we only need the linear part of the neural network.

Moreover, to accelerate its convergence, we utilize the channel hardening phenomenon \cite{channelhardening2014} to initialize the input vector ${{{\bf{\hat x}}}^q}_0$ as
\begin{equation}\label{eq:input}
{{{\bf{\hat x}}}^q}_0 = {{\bf{D}}^{ - 1}}{{\bf{H}}^T}{\bf{y}}.
\end{equation}

\subsection{Training Procedure}
The number of parameters in DNN is $2L$ in total, i.e., $\left\{ {\alpha _t^{\left( 1 \right)},\alpha _t^{\left( 2 \right)},t = 0, \cdots ,L-1} \right\}$. Note that $\alpha _0^{\left( 1 \right)}$ and $\alpha _0^{\left( 2 \right)}$ are the initial values and need not to be trained while the rest of the parameters $\left\{ {\alpha _t^{\left( 1 \right)},\alpha _t^{\left( 2 \right)},t = 1, \cdots ,L-1} \right\}$ are obtained through the training phase.

Specifically, since the residual error vector ${{\bf{v}}_0}$ has no obvious physical meaning, the coefficient $\alpha _0^{\left( 1 \right)}$ of ${{\bf{v}}_0}$ is initialized to be close to 0.
Meanwhile, the residual vector ${{\bf{v}}_0}$ is random but close to 0, which ensures that
the recursive accumulation of the residual vector could not surpass the accurate interpolation ${{\bf{v}}_r} = {\bf{x}} - {{{\bf{\hat x}}}_0}$ and guarantee the randomness.

Moreover, due to that ${{{\bf{\hat x}}}_0}$ obtained from \eqref{eq:input} achieves reasonable performance in massive MIMO scenario, the parameter $\alpha _0^{\left( 2 \right)}$ is set to be large.
Nevertheless, the large $\alpha _0^{\left( 2 \right)}$ would lower the randomness of the proposed algorithm.
Hence, $\alpha _0^{\left( 2 \right)}$ is set to be 0.5 to maintain the convergence and randomness.

In the training phase, loss function and optimization function, are utilized to adapt the parameters $\left\{ {\alpha _t^{\left( 1 \right)},\alpha _t^{\left( 2 \right)},t = 1, \cdots ,L-1} \right\}$ of the overall DNN for accurate detection.

The mean squared error (MSE) is adopted as the loss function to express the distance between the output of neural network ${{{\bf{\hat x}}}_L}$ and the transmitted vector $\mathbf{x}$ as
\begin{equation}\label{eq:11}
L\left( {{\bf{x}},{{{\bf{\hat x}}}_L}} \right) = \frac{1}{{2K}}\sum\limits_{i = 1}^{2K} {{{\left( {{x}\left( i \right) - {{{{\hat x}}}_L}\left( i \right)} \right)}^2}}.
\end{equation}

The adaptive moment estimation (ADAM) is utilized as the optimization function to minimize the loss function in \eqref{eq:11} and decide the learned parameters $\left\{ {\alpha _t^{\left( 1 \right)},\alpha _t^{\left( 2 \right)},t = 1, \cdots ,L-1} \right\}$ through the feedforward network.

The difference between the proposed algorithm and the iterative algorithm in \cite{Mandloi2017Low-Complexity} lies in the residual vector of previous layer and the convex combination, which multiply coefficients and then are added at each layer.
The so-derived parameters $\left\{ {\alpha _t^{\left( 1 \right)},\alpha _t^{\left( 2 \right)},t = 1, \cdots ,L-1} \right\}$ are learned by DL.
%

\subsection{Complexity Analysis}
In this subsection, the flops of the multiplication operation of the proposed algorithm  are compared with other algorithm, i.e., LMMSE \cite{Yang2015Fifty}, DetNet \cite{samuel2017Deep}\cite{samuel2019Learning} and the iterative algorithm in \cite{Mandloi2017Low-Complexity}.
\begin{table}[]
\centering
\caption{Complexity comparison of the detection algorithms}
\begin{tabular}{cc}
\hline
MIMO Detector & \multicolumn{1}{c}{Number of Flops Operation}                               \\ \hline
LMMSE          & ${K^3} + {K^2}$                                                             \\
DetNet        & $\left( {K\left( {128K - 2} \right)} \right){L}$                              \\
The iterative algorithm in \cite{Mandloi2017Low-Complexity}      & $4L{K^2} + 2\left( {2L + 1} \right)K$                   \\
Proposed      & $4L{K^2} + 2\left( {2L + 1} \right)K + 3KL $             \\ \hline
\end{tabular}
\label{tab:complexity}
\end{table}

As introduced in \cite{Tan2019ALow-Complexity}, the complexity of LMMSE is of order $\mathcal{O}({K^3} + {K^2})$.
However, the LMMSE detector needs to do the matrix inversion operation that is forbiddingly high with increasing antennas.
The number of operations for DetNet is $\mathcal{O}(\left( {K\left( {128K - 2} \right)} \right){L})$.
Besides, the work in \cite{Mandloi2017Low-Complexity} takes the $\mathcal{O}(4L{K^2} + 2\left( {2L + 1} \right)K)$ operations to detect the optimal signals,
which coincides with the previous part of the proposed algorithm.
Hence, the overall complexity of the proposed algorithm sums up to $\mathcal{O}(4L{K^2} + 2\left( {2L + 1} \right)K + 3KL) $ that adds $3KL$ based on the original work due to the two parameters multiplying three $K$ dimensional vector at each iteration.
The complexity comparison are presented in Tab. \ref{tab:complexity}.

Hence, the computational cost of the proposed algorithm is lower than that of LMMSE and DetNet.
Compared with the iterative algorithm \cite{Mandloi2017Low-Complexity},
the proposed one reduces BER.



\section{Simulation Results}

In this section, experiments are conducted to demonstrate the performance of the proposed DL-based detection in massive MIMO scenarios.
We compare with DetNet\cite{samuel2017Deep}, \cite{samuel2019Learning}, LMMSE\cite{Yang2015Fifty} and iterative algorithm in \cite{Mandloi2017Low-Complexity}.
%
%
%

\subsection{Dataset}
The settings of the DL-based detector are given in Tab. \ref{tab:settings}.
The training data is generated by transmitting random QPSK and 16QAM sequences through the additive white Gaussian channel.
The SNR is uniformly distributed on $\mathcal{U}\left( { - 1,21} \right)$ at the training stage and ranges from 0 dB to 13 dB at the testing stage.

\begin{table}[]
\centering
\caption{Training settings in numerical tests}
\begin{tabular}{c|p{2.65cm}<{\centering}|p{1.95cm}<{\centering}}
\hline
\hline
Parameters             & \multicolumn{2}{c}{Values}                                    \\ \hline
$N$                     & 128                 & 64                               \\\hline
$K$                      & 8                   & 8                                  \\\hline
Layers                 & \multicolumn{2}{c}{8}                                        \\\hline
SNRs for training      & \multicolumn{2}{c}{{[}-1,21{]}}                              \\\hline
Starting learning rate & \multicolumn{2}{c}{0.0001}                                   \\\hline
Mini-batch size        & \multicolumn{2}{c}{5000}                                     \\\hline
Size of training data  & \multicolumn{2}{c}{20000}                                    \\\hline
Activation function    & \multicolumn{2}{c}{Leaky ReLu for the first and Tanh for the second} \\\hline
Optimization method    & \multicolumn{2}{c}{ADAM optimizer}                           \\\hline\hline
\end{tabular}
\label{tab:settings}
\end{table}
%

The learning rate of the ADAM optimizer is ${\beta _0}=0.0001$ and decays exponentially after each epoch of training.
The learning rate declines after each epoch as
\begin{equation}\label{eq:15}
{\beta _t} = {\beta _0} \times {0.9^t},
\end{equation}
where ${\beta _t}$ is the learning rate after the $t$-th epoch.


\subsection{Performance of MIMO detector}
\begin{figure}[!t]
  \centering
  \includegraphics[width=8.7cm]{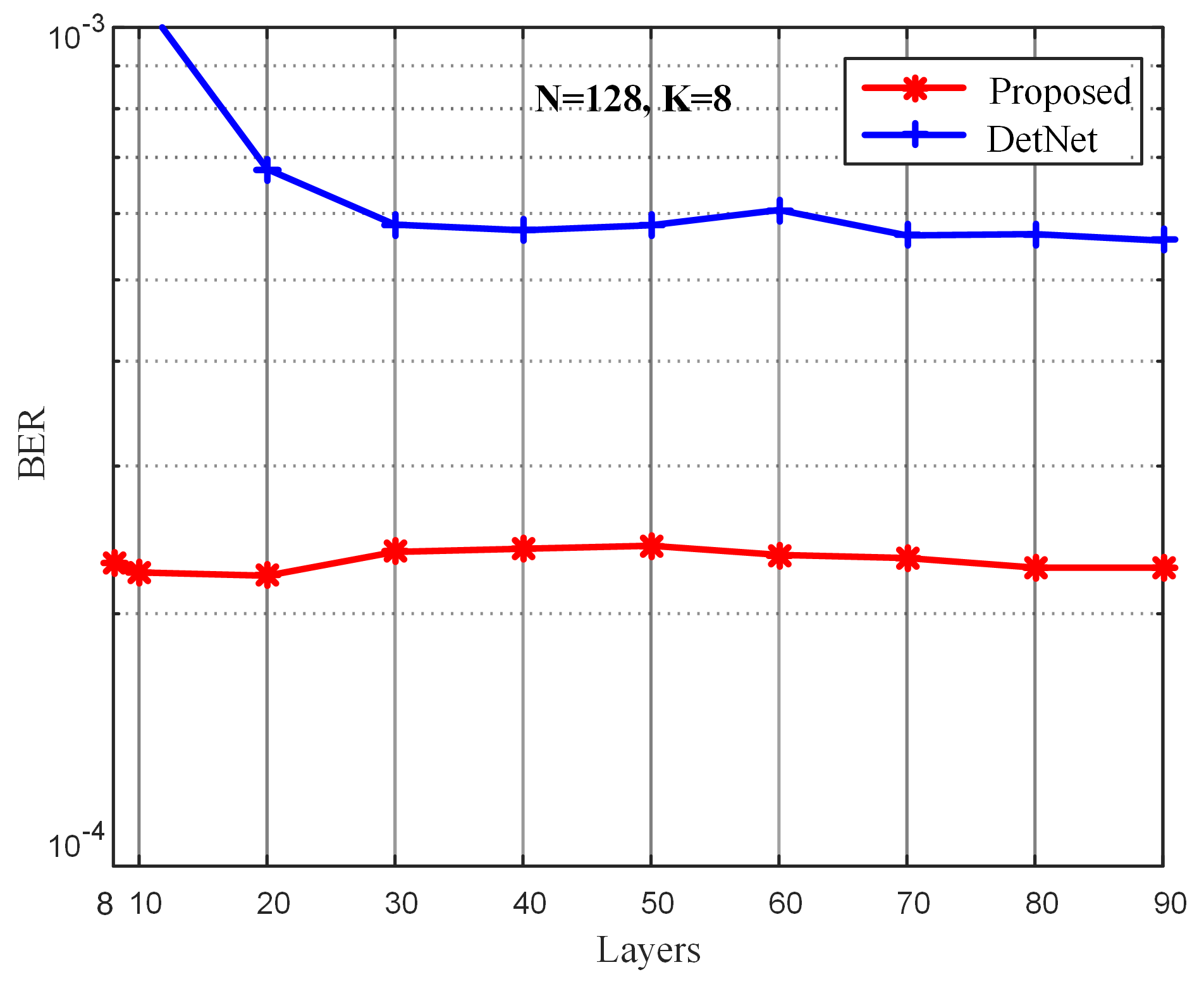}\\  
  \caption{Different DNN layers comparison for $128 \times 8$ antenna configuration at $\rm{SNR}=11dB$}
\end{figure}

\begin{figure}[!t]
  \centering
  \includegraphics[width=9cm]{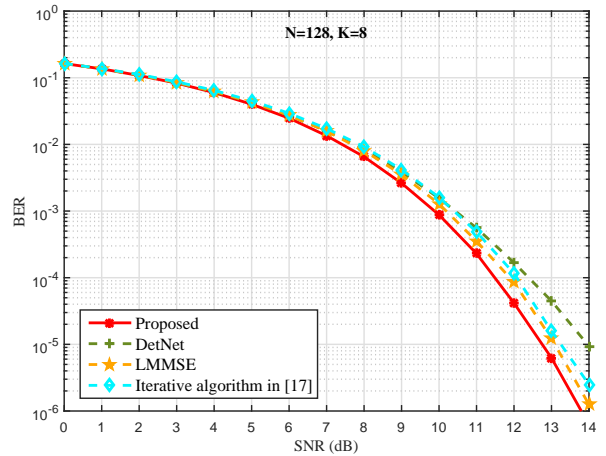}\\  
  \caption{BER curves of deep learning detectors using QPSK over the massive MIMO Rayleigh fading channel with $128 \times 8$ antenna configuration}
\end{figure}
Fig. 3 compares the performance of the proposed algorithm utilizing different layers with DetNet when the configuration is $128 \times 8$ for $\rm{SNR}=11dB$ and QPSK signals.
From the figure, the performance of DetNet improves as the number of layers grows, whereas the performance of the  proposed algorithm keeps stable.
Specifically, the proposed algorithm achieves better performance with as small as 8 layers while DetNet converges only after as many as 90 layers, which indicates the lower complexity by implementing the proposed method.
Hence, we choose $L = 8$ for the proposed algorithm and adopt $L = 90$ for DetNet in the following experiments.
Although we choose $L = 90$ for DetNet and $L = 8$ for the proposed algorithm, the performance of the proposed algorithm is still superior than DetNet.

Fig. 4 demonstrates that the performances of the four approaches detecting QPSK signals are similar at low SNR when the antenna configuration is $128 \times 8$ in each realization.
As the SNR grows,
the performance of the iterative algorithm in \cite{Mandloi2017Low-Complexity} and the DetNet are almost the same.
Meanwhile, the proposed method is superior to all other detectors and the LMMSE is the second best to the proposed method.
Obviously, the proposed detector shows more remarkable performance in BER as the SNR grows and when the number of antenna is large.

Fig. 5 shows the BER performance under QPSK modulation when the number of antennas reduces to $64 \times 8$ massive MIMO scenario.
\begin{figure}[!t]
  \centering
  \includegraphics[width=9cm]{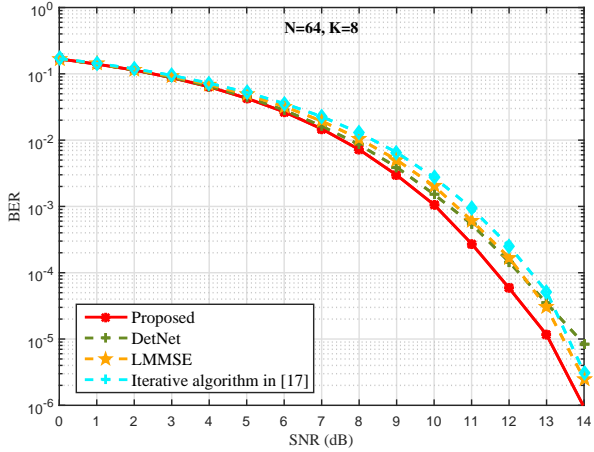}\\  
  \caption{BER curves of deep learning detectors using QPSK over the massive MIMO Rayleigh fading channel with $64 \times 8$ antenna configuration}
\end{figure}
In this case, the DL-based approaches have much better performance than the two traditional approaches including the iterative algorithm in \cite{Mandloi2017Low-Complexity} and the LMMSE method.
The DetNet outperforms the LMMSE method and the iterative algorithm in \cite{Mandloi2017Low-Complexity}, and approaches to the proposed method.
Nevertheless, the computational complexity of DetNet with 90 layers are much higher than that of the proposed method.
\begin{figure}[!t]
  \centering
  \includegraphics[width=9cm]{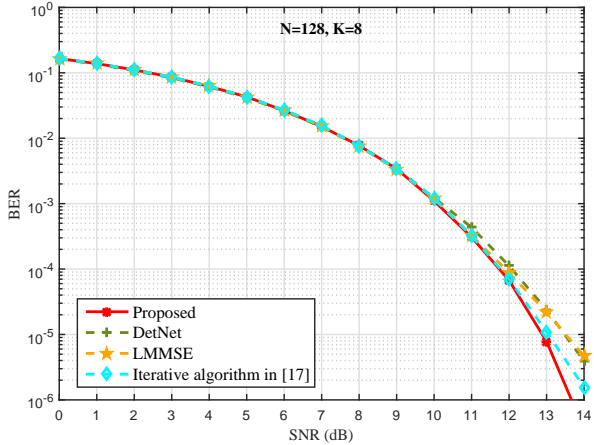}\\  
  \caption{BER curves of deep learning detectors using 16QAM over the massive MIMO Rayleigh fading channel with $128 \times 8$ antenna configuration}
\end{figure}

Lastly, Fig. 6 shows the BER performance for $128 \times 8$ massive MIMO scenario with 16QAM modulation.
Since DetNet in \cite{samuel2017Deep} is not suitable for high-order modulation situation (high BER), we use DetNet in \cite{samuel2019Learning} for comparison.
From this figure, the proposed algorithm still works as the best one among all algorithms.
The iterative algorithm in \cite{Mandloi2017Low-Complexity} is comparable to the proposed one.
DetNet in \cite{samuel2019Learning} has similar performance with LMMSE, but needs a long time to implement.
Moreover, the proposed algorithm has lower gain when compared to QPSK modulation Fig. 4 under the same antenna configuration.


\section{Conclusion}
In this paper, we have proposed a model-driven DL-based approach that is formulated by interference cancellation for massive MIMO scenarios.
The proposed algorithm is inversion-free, and therefore is computationally inexpensive.
The experiment has been conducted in various channel scenarios and indicates that the performance of the proposed method is superior than the existing detectors without any knowledge regarding the SNR level. 

\vspace{12pt}
\color{red}

\end{document}